\documentclass[showpacs,preprintnumbers,amsmath,amssymb]{revtex4}

\newcommand{\ds}{\displaystyle}

\newcommand{\be}{\begin{equation}}
\newcommand{\en}{\end{equation}}
\newcommand{\bea}{\begin{eqnarray}}
\newcommand{\ena}{\end{eqnarray}}

\usepackage{graphicx}
\usepackage{bm}

\begin{document}

\baselineskip 24pt

\title{Tachyonic open inflationary universes }
\author{Leonardo Balart, Sergio del Campo, Ram\'on Herrera,
Pedro Labra{\~n}a and Joel Saavedra}
\address{Instituto de F\'{\i}sica, Pontificia Universidad Cat\'olica de
Valpara\'{\i}so, Casilla 4950, Valpara\'{\i}so}

\date{\today}

\begin{abstract}
We study one-field open inflationary models in a universe
dominated by  tachyon matter. In these scenarios, we determine and
characterize the existence of the Coleman-De Lucia (CDL)
instanton. Also, we study the Lorentzian regime, that is, the
period of inflation after tunnelling has occurred.

\end{abstract}

\pacs{98.80.Jk, 98.80.Bp}
\maketitle

\section{\label{Int} Introduction}

Cosmological inflation has become an integral part of the standard
model of the universe. Apart from being capable of removing
the shortcomings of the standard cosmology, it gives important clues
for structure formation in the universe. The scheme of inflation \cite{IC}
(see \cite{libro} for a review) is based on the idea that
at early times
there was a phase in which the universe evolved through
accelerated expansion in a short period of time at high energy
scales. During this phase, the universe was dominated by a
potential $V(\phi)$ of a scalar field $\phi$, which is called the
inflaton.

Normally, inflation has been associated
with a flat universe, due to
its ability to effectively drive the spatial
curvature  to zero. In fact, requiring sufficient
inflation to homogenize random initial conditions, drives the
universe very close to its critical density.
In this context, the recent observations from WMAP~\cite{wmap} are
entirely consistent with a universe having a
total energy density that is very close to its critical value,
where the total density parameter has the value $\Omega =1.02\pm
0.04$. Most people interpret this value as corresponding to a flat
universe. But, according to this result, we might take the
alternate point of view of a marginally open or closed
universe model~\cite{ellis-k}, which at early times in the evolution
of the universe presents an
inflationary period of expansion. This approach has already been
considered in the literature~\cite{linde, re8, shiba, delC1, delC2}
in the context of the Einstein theory of
relativity, Jordan-Brans-Dicke or brane world cosmology.
At this point, we should mention that
Ratra and Peebles were the first to elaborate on the open
inflation model \cite{Ratra}.
The basic idea in an open universe is that a symmetric bubble
nucleates in the de Sitter space background, and its interior
undergoes a stage of slow-roll inflation, where the parameter
$\Omega_0$ can be adjusted to any value in the range $0 < \Omega
_0 < 1$.
Bubble formation in the false vacuum is described by the
Coleman-De Lucia(CDL) instantons~\cite{CDL}. Once a bubble has
taken place by this mechanism, the inside of the bubble
looks like an
infinite open universe. The problem with this sort of scenario is
that the instanton exists only if the following inequality $\mid
V,_{\phi\phi}\mid > H^{2}$ is satisfied during part of
the tunneling process  \cite{Hawking:1981fz}. On
the contrary, during inflation the inequality $\mid V,_{\phi\phi}\mid \ll
H^{2}$ is satisfied (slow-roll approximation).
Linde solves this problem by proposing a simple one-field model in
Einstein's general relativity (GR) theory~\cite{linde, re8} where the crucial
point is the very peculiar shape of the effective scalar potential.
This scheme has been also successfully used in other more general
models \cite{delC1,delC2}.

On the other hand, as was mentioned before, one normally considers the
inflation phase driven by the potential or vacuum energy of a
scalar field, whose dynamics is determined by the Klein-Gordon
action. However, more recently and motivated by string theory,
other non-standard scalar field actions have been used in
cosmology. In this context the deep interplay between small-scale
non-perturbative string theory and large-scale brane-world
scenarios has raised the interest in a tachyon field as an
inflationary mechanism, especially in the Dirac-Born-Infeld action
formulation as a description of the D-brane action \cite{DBI}.
In this scheme, rolling tachyon matter is associate with
unstable D-brane. The decay of these D-branes produce a pressureless
gas with finite energy density that resembles classical dust.
Cosmological implications of this rolling tachyon were first studied
by Gibbons \cite{Gibbons:2002md} and in this context it is
quite natural to consider
scenarios where inflation is driven by this rolling tachyon.

From the above point of view, we shall study a single-field open
inflationary universe, where inflation is driven by a tachyon
field. Therefore, we model our potential by a single tachyon field
$\phi$ which presents an exponential term and a barrier generated
by a expression which presents a peak with a given maximum (see
Fig.\ref{V}). The barrier term controls the bubble nucleation and
the exponential term controls inflation after quantum tunnelling.
In particular, in our model the tachyon begins at the false
vacuum and then decays to the value $\phi_T$ via quantum tunnelling
through the potential barrier (see Fig.\ref{V}),
generating during this process an open inflationary universe.
The mechanism of tachyon decay via quantum tunnelling has been
studied in the context of brane-antibrane and dielectric brane
decays \cite{Hashimoto:2002xt}, where tachyon potential with similar
characteristics of our potential have been derived and
studied.
In our work we are particulary interested in the cosmological
implication of this process and its application to open
inflationary models.

The paper is organized as follows. In Sec.~\ref{Sec1}, we write
and numerically solve the tachyon field equations in an Euclidean
space-time. Here the existence of the CDL instanton for a tachyon
model is studied, together with the basic properties of the bubble
which is created during the tunnelling process. In
Sec.~\ref{Sec2}, we study the characteristics of the open
inflationary universe model that is produced after tunnelling has
occurred. We determine the corresponding scalar density
perturbations and the scalar and tensorial spectral indexes. We
summarize our points and give conclusions in Sec.~\ref{Sec3}.

\section{\label{Sec1}The Euclidean cosmological equations in tachyon models}

We consider the Euclidean effective action of the tachyon field
given by \cite{sen}
\begin{eqnarray}
S \, = \, \int{d^{4}x\,\sqrt{-g}}\,\left
[\,\frac{1}{2\kappa}\,\,R\, +V(\phi)\,\sqrt{1+\partial_{\mu}\phi
\partial^{\mu}\phi}\,\right], \label{ac1}
 \end{eqnarray}
where $R$ is the Ricci scalar curvature, $\kappa = 8\pi G$, the cosmological
constant term is set to zero and $V(\phi)$ is an effective scalar
potential associated with the tachyon field $\phi$ (with unit
 $\kappa^{1/2}$) given by
\be \label{pot} V(\phi)=V_0\,e^{-\lambda
\phi}\left[1+\frac{\alpha^2}{\beta^2+(\phi - v)^2}\right]. \en

\begin{figure}[t]
\includegraphics[width=3.2in,angle=0,clip=true]{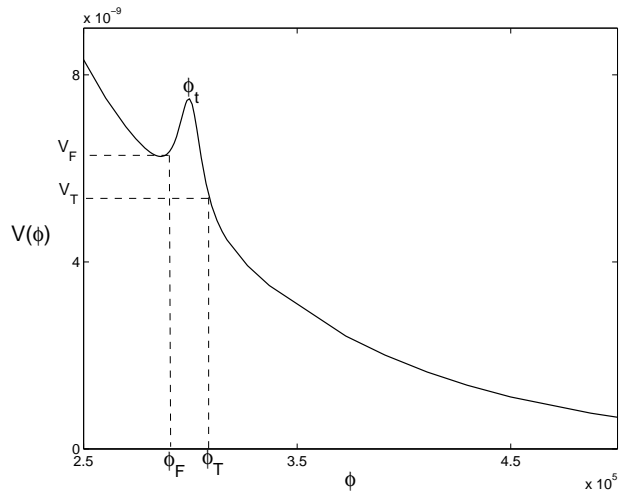}
\caption{Scalar potential associated to the tachyon field. All
values are given in units where $\kappa =1$. In the graph,
some relevant values which the tachyon field gets during
the process are schematically indicated.
They are the false vacuum $\phi_F$
from which the tunnelling begins, $\phi_T$ where the tunnelling
stops and where the inflationary era begins, while $\phi_t$
denote the top of the barrier.} \label{V}
\end{figure}

We are going to consider $\lambda$  and $V_0$  as free parameters,
and $\alpha$, $\beta$ and $v$ as arbitrary constants, which will
be set by phenomenological considerations, as we shall show below.
In the following, we shall take $\lambda>0$. The first term of the
effective potential controls inflation after quantum tunnelling
has occurred. Its form coincides with that used in the simplest
tachyonic inflationary universe model where $V(\phi) = V_0
e^{-\lambda\phi}$ \cite{sami}. In the same spirit of Ref.
\cite{linde}, the second term controls the bubble nucleation,
whose role is to create an appropriate shape in the tachyon
potential, $V(\phi)$, where its local maximum occurs at the top of
the barrier $\phi = \phi_t$ (see Fig.\ref{V}). After the quantum
tunnelling, we should note that the stable vacuum to which the
tachyon condenses, is at $\phi \rightarrow +\infty$, when
$V(\phi)\rightarrow 0$.

Tachyon models which are classically stable but decay
by quantum tunnelling have been used in the context of non-standard
D-brane decays. In these models it has been reported that together
with normal
annihilation of branes via gravitational forces, branes can decay in
another way, that is through the tunnel effect by creation of a throat between
the brane and the antibrane
\cite{Hashimoto:2002xt,Callan:1997kz,Savvidy:1998xx}.
In these models, tachyon potentials with similar characteristic as
(\ref{pot}) appear, due for example, to the $U(1)$ charge of the
tachyon field in the brane-antibrane configuration.

Nevertheless our model has been motivated by string theory,
%
%
we consider it just at a
phenomenological level without claiming any
direct identification of $\phi$ with the string tachyon
field. Indeed, as it was mentioned in
Refs.~\cite{Frolov:2002rr,Kofman:2002rh},
there are problems with inflation paradigm when the origin of
$\phi$ is traced in string theory. Then, we take
the approach of Refs.~\cite{sami,ste-ver,Sami:2003qx} and
set the parameters of the model according to phenomenological
consideration.

The $O(4)$- invariant Euclidean space-time metric is described by
\be \ds d{s}^{2}\,=\, d{\tau}^{2}\,+\, a^2(\tau)\, \left[\,\,
d{\psi}^{2}\,+\,\sin^{2}(\psi)\,d{\Omega^{2}_{2}}\,\,\right],
\label{met} \en where $a(\tau)$ is the scale factor of the
universe and $\tau$ represents the Euclidean time.

The equation governing the evolution of factor $a(\tau)$ is
 \be \ds
\left(\frac{a'}{a}\right)^{2}
=\frac{1}{a^{2}}-\frac{\kappa}{3}\left[\frac
{V(\phi)}{\sqrt{1+\phi\, '^2}} \right]\, \label{ec2},
 \en
and, after varying~(\ref{ac1}) with respect to $\phi$, we obtain
the equation of motion of the tachyon field \be \ds
\frac{\phi\,''}{1+\phi\,
'^2}\,=\,-3\,\frac{a'}{a}\,\phi\,'\,+\,\frac{1}{V}\,
\frac{dV}{d\phi}\,\,\label{ec4}, \en where the primes denote
derivatives with respect to $\tau$.

 From Equations~(\ref{ec2}) and~(\ref{ec4}) we obtain

\begin{equation}
\label{ec5}
a''= - \frac{\kappa}{3}\,a\,\frac{V(\phi)}{\sqrt{1+{\phi'}^2}}
\left( 1+\frac{3}{2}\,{\phi'}^2 \right).
\end{equation}
%
%
Now, we proceed to find a criterion for the existence of a CDL
instanton in a tachyon model. This criterion could be found in a
similar way as it is done in an universe described by Einstein's
General Relativity, where the matter content is described by a
scalar field \cite{linde}. In this approach, the bubbles have size
greater than the Compton wavelength of the scalar field, $r\gtrsim
m^{-1} \sim (V_{, \,\phi \; \phi})^{-1/2}$, where $m$ is the mass
associated of the scalar field. In this way, the instanton CDL
solution can exist only if the bubble can fit into de Sitter
sphere of radius $r \sim H^{-1}$, therefore the CDL instantons are
possible only if $H^{2}<V_{, \,\phi \; \phi}$. In our case,
according to reference \cite{delaMacorra:2006tm} the mass of the
tachyon field is given by
\begin{equation}
m^2=\frac{V_{, \,\phi \;
\phi}}{V}-\left(\frac{V_{,\,\phi}}{V}\right)^2\;\left(3-\frac{V^2}{\rho^2}
\right),
\end{equation}
where $\rho$ is the  energy density of the tachyon field.
Therefore, the CDL instantons are possible in the tachyonic case
if the  condition
\begin{equation}
m^2=\frac{V_{,\, \phi \;
\phi}}{V}-\left(\frac{V_{,\,\phi}}{V}\right)^2\;\left(3-\frac{V^2}{\rho^2}
\right)>\;H^2,\label{criterio}
\end{equation}
is satisfied during the tunnelling process.

Now, in order to obtain a numerical solution of the field
equations (\ref{ec4}) and (\ref{ec5}), we will choose a particular
election of the parameters that appears in the scalar potential
(\ref{pot}). In particular, we use the COBE normalized value for
the amplitude of the scalar density perturbations in order to
estimate $\lambda$ and $V_0$ \cite{sami}. Thus we have $\lambda =
10^{-5}\kappa^{-1/2}$ and $V_0 = 10^{-7} \kappa^{-2}$. We also
consider $\beta^{2} = 2 \alpha^{2}$ with $\beta = 6.67\times
10^3\,\kappa^{1/2}$ and $v = 3\times 10^5\,\kappa^{1/2}$, which,
as we will see, will provide about 46 e-folds of inflation after
the tunnelling. The lower value of the e-folding is not a problem,
since in the context of the tachyonic-curvaton reheating, it could
be of the order $45$ or $50$, since the inflationary scale can be
lower \cite{Campuzano:2005qw}.

%
\begin{figure}[t]
\includegraphics[width=3in,angle=0,clip=true]{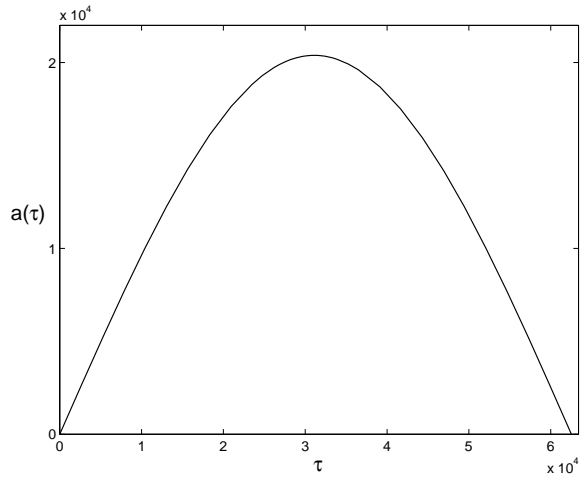}
\caption{This graph shows the plot of the scale factor $a(\tau)$
as a function of the Euclidean time $\tau$.} \label{a1}
\end{figure}
%

%
\begin{figure}[t]
\includegraphics[width=3in,angle=0,clip=true]{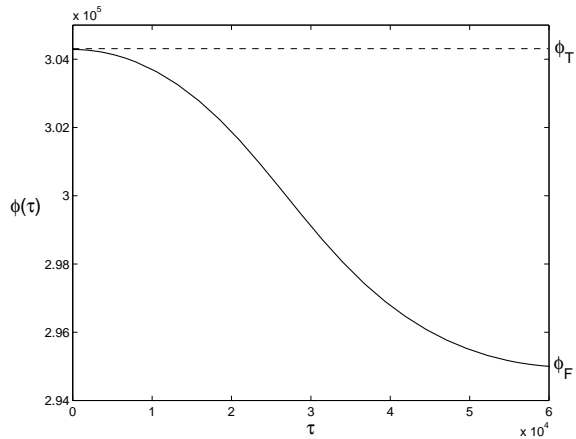}
\caption{Numerical solution of Eq.~(\ref{ec4}) corresponding
to the Coleman-De Lucia
instanton in our model.} \label{taq1}
\end{figure}
%

\begin{figure}[t]
\includegraphics[width=3in,angle=0,clip=true]{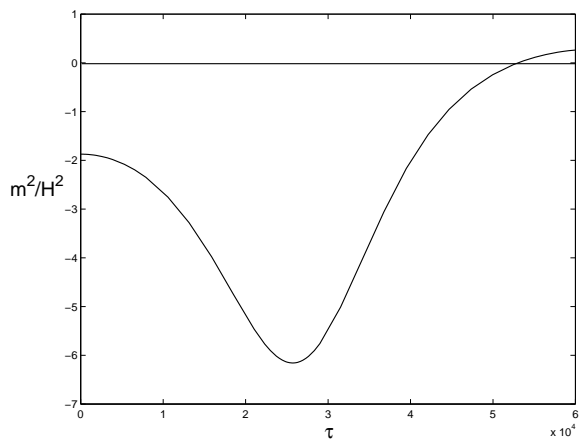}
\caption{This plot shows how during the tunneling process the
inequality $m^2> H^{2}$ is hold.} \label{cod}
\end{figure}

We solve the Equations (\ref{ec2}) and (\ref{ec5}) considering the
following boundary conditions: $\phi = \phi_T$, $\phi'= 0$, $a =
0$, $a'= 1$, at $\tau = 0$ and working in units where $\kappa=1$.
Actually, the instanton has the topology of a four-sphere and
there are two places where the scale factor $a(\tau)$ vanishes.
These are the points at which $\tau=0$ and $\tau=\tau_{max}$.
Then, the boundary conditions on $\phi$ arise from the requirement
that the term in the scalar field Eq.(\ref{ec4})
$3\phi^{'}a^{'}/a$, be finite at these point i.e.,
$\phi^{'}(0)=\phi^{'}(\tau_{max})=0$. From Eq.(\ref{ec2}), we get
that at the points where of the scalar factor vanishes we have,
$a^{'}=\pm 1$. Figure \ref{a1} shows how the scale factor evolves
during the tunneling process. A numerical solution which
corresponds to the CDL instanton is showed in Fig.\ref{taq1}. In
that case tunnelling occurs from $\phi_F\approx 2.95\times 10^5$
to $\phi_T\approx 3.043\times 10^5$. Notice that the tachyon
potential considered in our model satisfies the criterion for the
existence of the instanton, i.e. satisfies the inequality
(\ref{criterio}). In order to see this, in Fig.~(\ref{cod}) we
have plotted $m^2\,/\,H^{2}$ as a function of the Euclidean time
$\tau $ for our model. From this plot we observe that most of the
time during the tunneling, we obtain $|m^2|\,>\,H^{2}$.

In the following, we calculate the instanton action
for the quantum tunnelling between the values $\phi_F$ and
$\phi_T$ in the effective tachyon potential, see Fig.~(\ref{V}).
By integrating by parts and using the
Euclidean equations of motion, we find that the action may be
written as

\begin{eqnarray}
\label{action111}
 S = 4\pi^{2}\,\int\,d\tau\left[a^{3}V(\phi)
 \frac{(1+\phi\,'^2/2)}{\sqrt{1+\phi\,'^2}}-\frac{3}{\kappa}
 a\right]
= \frac{12\pi^{2}}{\kappa}\,\int\,d\tau\left[\frac{1}{2}a\;\phi\,'^2-
 a a\,'^2\left(1+\frac{\phi\,'^2}{2}\right)\right].
\end{eqnarray}

The tachyon field $\phi$ is initially trapped in its false vacuum,
whose value is $\phi_{F}$. After tunnelling
the instanton gets the value $\phi_{T}$ and a single bubble is
produced. The bounce-action is given by $B = S_B - S_{F}$, i.e.
the difference between the action associated with the bounce
solution and the false vacuum. This action determines the
probability of tunnelling for the process. Under the approximation
that the bubble wall is infinitesimally thin, we obtain a first
integral of Eq. (\ref{ec4}) given by \be
\phi\,'\,^2=\left(\frac{V(\phi)}{V_T}\right)^2 -1 ,\en
where we have defined $V_{T}=V(\phi_{T})$.

Outside the wall, $\phi = \phi_F$, then we can write
$$
B_{outside}=0.
$$

Within the wall, $a=R_b$ where $R_b$ is the radius of the bubble
and after using the thin-wall approximation, we get
\be
B_{wall}=\frac{6\pi^2}{\kappa}\,R_b\,S_1\,,
 \en
where the surface tension of the wall becomes defined by $S_1$ and
is given by
\be S_{1}\, =\,\int_{\phi_{T}}^{\phi_{F}}d\phi\left[\left(\frac{V(\phi)}{V_{T}}
\right)^2-1\right]^{1/2}
 \en

Inside the wall $\phi = \phi_T$, and we obtain
\begin{equation}
B_{inside}=\frac{12\pi^2}{\kappa^2}\left(\frac{1}{V_T}\left[
(1-\kappa V_T R_b^2/3)^{3/2}-1 \right] -
\frac{1}{V_F}\left[
(1-\kappa V_F R_b^2/3)^{3/2}-1 \right] \right),
\end{equation}
where we have defined $V_F=V(\phi_F)$.
Thus, the total reduced bounce-action for the thin-wall bubble results to be:
\begin{equation}
\label{a10}
 B=\frac{6\pi^2}{\kappa}\,R\,S_1 +
 \frac{12\pi^2}{\kappa^2}\left(\frac{1}{V_T}\left[
(1-\kappa V_T R_b^2/3)^{3/2}-1 \right] -
\frac{1}{V_F}\left[
(1-\kappa V_F R_b^2/3)^{3/2}-1 \right] \right),
\end{equation}
where, we have taken into account the contributions from the wall
(first term) and the interior of the bubble (the second and third
terms).

The radius of curvature of the bubble
could be obtained demanding that
the bounce-action~(\ref{a10}) is an extremum. Then, the
radius of the bubble is determined by setting $dB/dR_b$ = 0, which
gives
\begin{equation}
\label{ecradio}
 \frac{S_{1}}{2\,R_b}=\left[\left(1-\frac{\kappa V_{T}R_b^{2}}{3}\right)^{1/2}-
 \left(1-\frac{\kappa V_{F}R_b^{2}}{3}\right)^{1/2}\right].
\end{equation}
Thus, the radius of the bubble, $R_b$, is found by solving this
equation. Also, from Eq.~(\ref{ecradio}) and following similar
arguments as those found in Ref.~\cite{MSTTYY}, we could obtain the following
condition for our configuration to describe the
$O(4)$-symmetric false vacuum decay
 \be \ds
\Delta\,s=\,\frac{S_{1}}{2\,R_b}\,<\,1, \label{bcond}
 \en
where, the dimensionless quantity, $\Delta\,s$, represents the
strength of the wall tension in the thin-wall approximation in our model.

The
radius of the bubble associated to the numerical solution
displayed in Fig.\ref{taq1}, could be found
from Equation (\ref{ecradio}) yielding $R_b =664.547$.
We also can determinate the strength of the
wall tension, $\Delta\,s = 3.302 \times 10^{-5}$, result
that satisfies the consistency condition for the false vacuum
decay Eq. (\ref{bcond}).



\section{\label{Sec2}Inflation after tunnelling and scalar perturbation spectra}

Let us study what happens after the tunnelling. In
order to do that we make an analytical continuation to the
Lorentzian space-time and study the time evolution of the tachyon
field $\phi(t)$ and scale factor $a(t)$. The equations of motion
are:
\be \ds \frac{\ddot{\phi}}{1-\dot{\phi}^2}\,=\,-3\,\frac{\dot{a}}{a}\,
\dot{\phi}\,-\,\frac{1}{V}\,
\frac{dV}{d\phi}\,\,\label{ec4L}, \en
\begin{equation}
\ddot{a} = \frac{\kappa}{3}\,a\,\frac{V(\phi)}{\sqrt{1-\dot{\phi}^2}} \left(
1-\frac{3}{2}\,\dot{\phi}^2 \right),
\end{equation}
where the dots represents derivative whit respect to the cosmological time $t$.

\begin{figure}[t]
\includegraphics[width=3.2in,angle=0,clip=true]{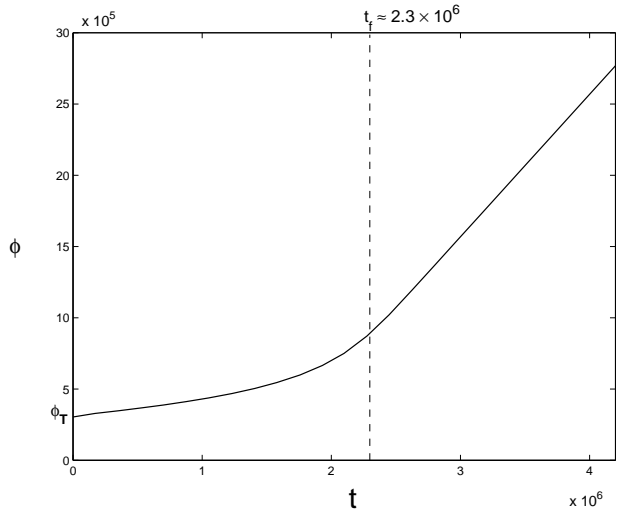}
\caption{The behavior of the tachyon field after the tunnelling
as a function of the cosmological time.
Notice that the different values
of $\phi$ are such that the slow-roll condition is satisfied during
inflation ($0 \leq t < t_f$). Also we can note that the scalar field
after inflation ($t \geq t_f$) does not oscillate.}
\label{phi1}
\end{figure}

\begin{figure}[ht]
\includegraphics[width=3.2in,angle=0,clip=true]{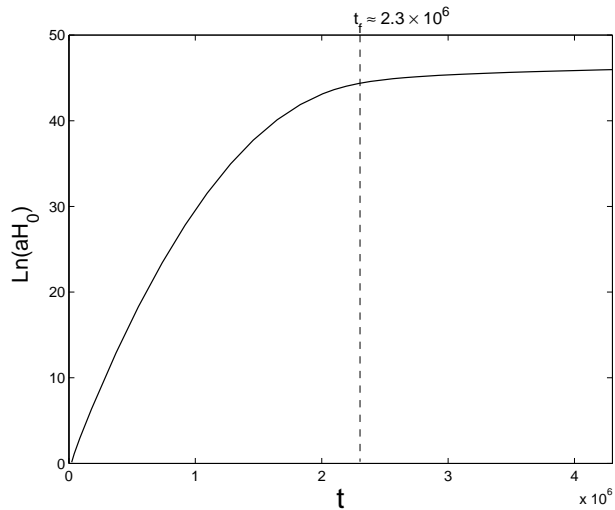}
\caption{The growth of the logarithm of the scale factor as a function
of the cosmological time. Here $H^2_0=V_T/3$.} \label{N1}
\end{figure}

These equations are solved numerically. This requires to
consider the parameters of the potential defined in the
previous section and the following boundary conditions $\phi(0)=\phi_T$,
$\dot{\phi}(0)=0$, $a(0) =0$, $\dot{a}(0)=1$.
We also consider units where $\kappa=1$. Solutions to these
equations are show in Fig. \ref{phi1} and Fig. \ref{N1}, where we
have taken $\phi_T
\approx 3.043\times 10^5$, a value obtained from the Euclidean numerical
solution.
We note in Fig. \ref{phi1} that after the tunnelling, during the period
$0\leq t < t_f$, the tachyon field satisfies the the slow roll
condition $\dot{\phi}^2 < 2/3$. Then, we conclude that after the
tunnelling our model generates a consistent inflationary era, where
the end of inflation happen at the cosmological time
$t = t_f$.
On the other hand, the scale factor expands
approximately $e^{46}$ times during this period, fact shown in
Fig. \ref{N1}.
%
%

Even though the study of scalar density perturbations in open
universes is quite complicated~\cite{re8}, it is interesting to
give an estimation of the standard quantum scalar field
fluctuations inside the bubble for our scenario.
In particular, the spectra of scalar perturbations for a flat space,
generated during tachyon
inflation, expressed in terms of the slow-roll parameters defined
in Ref. \cite{Schwarz:2001vv} becomes \cite{ste-ver}:
\be \frac{\delta\,\rho}{\rho}\, = \left[1 -
0.11\, \epsilon_1 + 0.36 \, \epsilon_2 \right]\frac{\kappa\,H}{2\pi\,\sqrt{2
\epsilon_1}}\,, \label{ec10}
 \en
where the slow-roll parameters are given by:
\begin{eqnarray}
 \epsilon_1 &\simeq&\frac{1}{2\,\kappa}\frac{(V\!,_{\phi})^2}{V^3}\,, \\
\epsilon_2 &\simeq& \kappa^{-1}\,\Big[-2\frac{V,_{\phi\phi}}{V^2}+ 3
\frac{(V\!,_{\phi})^2}{V^3}\Big].\label{e2}
 \end{eqnarray}

Certainly, in our case, Equation (\ref{ec10}) is an approximation
and must be supplemented by several
different contribution in the context of an open inflationary
universe \cite{linde}.
However, one may expect that
the flat-space expression gives a correct result for $N>3$.
%
\begin{figure}[t]
\includegraphics[width=3.2in,angle=0,clip=true]{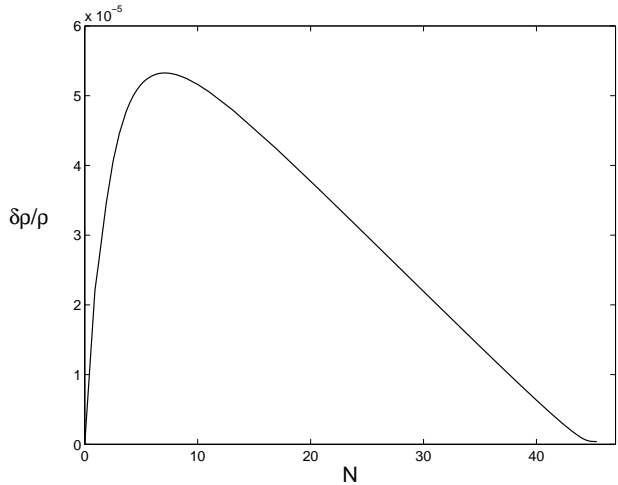}
\caption{Scalar density perturbations in our model produced inside
the bubble, $N$ e-folds after the open universe creation. Note
that our values coincide with the result of COBE normalized
value $\delta_H\!\sim\! 2\times10^{-5}$ \cite{COBE Nor}.}\label{Pert2}
\end{figure}
%
Figure \ref{Pert2} shows the magnitude of the scalar density
perturbations $\delta\rho/\rho$ for our model as a function of the
$N$ e-folds of inflation, after the open universe was formed.
Notice that $\delta\rho/\rho$ has a deep minimum at $N\lesssim 3$
(mechanism of suppression of large scale density perturbation)
and then approaches to its maximum at $N \sim 7$.
The shape of the graph showed in Fig. \ref{Pert2} seems to be a
generic feature of the
one-field open inflation models based on tunnelling and bubble formation
\cite{linde,delC1,delC2}.
Even though in our model
$\delta\rho/\rho$ has a maximum at small $N$, as opposed to the standard
single scalar inflationary models, where the maximum is located at
$N \sim 10$. If $N=O(1)$ corresponds to density perturbations on the
horizon scale $\sim 10^{28}$cm, then the maximum of the
spectrum appears on a scale which is about three orders of magnitude
smaller $\sim 10^{25}$cm, in our model.

One interesting parameter to consider is the so-called spectral
index $n$, which is related to the power spectrum of density
perturbations $P^{1/2}_{\cal R}(k)$. For modes with a wavelength
much larger than the horizon ($k \ll a H$), the spectral index $n$
is an exact power law, expressed by $P^{1/2}_{\cal R}(k) \propto
k^{n-1}$, where $k$ is the comoving wave number. Also it is interesting
to give an estimate of the tensor spectral index $n_T$.
In tachyon inflationary models the scalar spectral index
and the tensor spectral index are given by
$n=1-2\epsilon_1 - \epsilon_2$ and $n_T = -2\epsilon_1$, in the
slow-roll approximation \cite{ste-ver}. From the numerical
solution we can obtain their values. In particular for $N\sim 7$
we have $n \approx 0.98$ and $n_T \approx -0.033$. Notice that
those indices are very closed to the Harrison-Zel'dovich spectrum
\cite{Ha-Ze}.

\section{conclusion and Final Remarks}
\label{Sec3}

Since we still we do not know the exact value of the $\Omega$
parameter, it is convenient to count on an inflationary universe
model in which $\Omega < 1$. In this sense, we could have
single-bubble open inflationary universe models, which may be
consistent with a natural scenario for understanding the large
scale structure. However, open inflationary models have a more
complicated primordial spectrum than the one obtained in flat
universes. Here, extra discrete modes and possibly large tensor
anisotropies spectrum could be found, especially those related to
supercurvature modes, which are particular to open inflationary
universes. Forthcoming astronomical measurements will determine if
this extra terms are present in the scalar spectrum.
Here, we have studied a one-field open inflationary universe
model, where inflation is driven by a
tachyonic field.
In particular we focus on the viability of the model in order
to generate a consistent inflationary scenario, compatible with
observations (number of e-folding, density perturbations, spectral
and tensor spectral indices).
The idea of consider a tachyon field
to drive inflation is a natural choice since the tachyon is a
unstable particle.
%
On the other hand the possibility that this tachyon generates
an open
universe via bubble nucleation by quantum tunnelling
appears as an interesting possibility to explore, given that,
in the context
of nonstandard brane decays \cite{Hashimoto:2002xt},
tachyon potentials with similar
characteristics as our potential appear.



We have found that the generalized condition for the existence of
the CDL instanton in a tachyonic universe is satisfied together
with the requirement of tachyonic inflation (slow-roll condition).
In this sense, our models are different to that corresponding to
the scalar inflaton case \cite{linde}, where both conditions are
oppositive to each other. In this sense, we have provided a model
in which both conditions are satisfied simultaneously, and thus, a
slow-roll open inflationary tachyonic dominated universe could be
realized.

We have generalized the CDL instanton action to tachyonic
dominated universe. This action is described by expression
(\ref{action111}) and was used to study the probability of
nucleation of a bubble. We found that this probability differs
from the usual obtain in the context of open scalar inflation. We
determined the corresponding wall tension, the strength of the
wall tension and the radius of the bubble and found that these
results are in agrement with the consistency condition for the
false vacuum decay Eq.~(\ref{bcond}).

In addition, we computed what happens after the tunnelling. In order to
do this we solved numerically the Lorentzian field equation in our
theory. Essentially,
we found that after the tunnelling, during the period $0\leq t < t_f$
(see Fig. \ref{phi1}),
the tachyon field satisfies the slow roll condition
and then an inflationary era is realized.
On the other hand the scale factor
expands approximately $e^{46}$ times during this period. The lower
value of the e-folding is not a problem, since in the context of
the tachyonic-curvaton reheating, the e-folding could be of the
order $45$ or $50$, since the inflationary scale can be lower
\cite{Campuzano:2005qw}.

We gave an estimate of the scalar density perturbation, where we
found that the values calculated during the slow roll inflation
coincide with the result of COBE normalized value,
$\delta_H\!\sim\! 2\times10^{-5}$ \cite{COBE Nor}. Also we have to
mention that the indices computed in the tachyonic model are very
closed to the Harrison-Zel'dovich spectrum.

In this way, we have shown that one-field open inflationary
universe models can be realized in the tachyonic theory in the
context of the Standard Einstein General Relativity. We hope, that
the most natural study for tachyonic theory would be realized for
the Brane World Cosmology scenarios and we leave this task for a
near future.

Finally, in principle we could compare our model, i.e. the
tachyonic  model, with that refereed  to the standard case i.e.
the no-tachyon case. However, at the moment we must say that in
both cases a fine tuning of the parameters appearing in the scalar
potential is required in order to get a suitable model which may
describe an inflationary period for the evolution of the universe.
In fact, as far as we know, we could not say which of these models
need more fine tuning that  the other one. Perhaps when new
measurements of astronomical observations be available we will be
able to say which one of these models needs less fine tuning of
its parameters.

\begin{acknowledgments}
L. B. is supported from PUCV through Proyecto de Investigadores
J\'ovenes a\~{n}o 2006. S. del C. was supported from COMISION
NACIONAL DE CIENCIAS Y TECNOLOGIA through FONDECYT Grants
\mbox{N$^{0}$ 1030469}, \mbox{N$^{0}$ 1040624} and \mbox{N$^{0}$
1051086}, and  was also partially supported by PUCV Grant N$^0$
123.764. R. H. was supported by Programa Bicentenario de Ciencia y
Tecnolog\'{\i}a through Grant Inserci\'on de Investigadores
Postdoctorales en la Acade\-mia \mbox {N$^0$ PSD/06}. P. L. is
supported from COMISION NACIONAL DE CIENCIAS Y TECNOLOGIA through
\mbox{FONDECYT} Postdoctoral Grant N$^0$ 3060114. J. S. was from
COMISION NACIONAL DE CIENCIAS Y TECNOLOGIA through \mbox{FONDECYT}
Grant 11060515 and supported by PUCV Grant N$^0$123.785.
\end{acknowledgments}

\end{document}